\documentclass{optica-article}

\journal{opticajournal} 

\articletype{Research Article}

\usepackage{lineno}
\usepackage{bm}
\usepackage{braket}
\usepackage{amsfonts}
\usepackage{float}
\usepackage{lipsum}
\usepackage{fancyhdr}
\usepackage[utf8]{inputenc}

\usepackage{hyperref}   
\usepackage{soul}       

\def\CC{\mathbb{C}}

\newcommand{\cG}{\mathcal{G}}

\newcommand{\cO}{\mathcal{O}}

\newcommand{\be}{\begin{equation}}
	\newcommand{\ee}{\end{equation}}
\newcommand{\bea}{\begin{eqnarray}}
	\newcommand{\eea}{\end{eqnarray}}

\newcommand{\ed}{\end{document}}

\newcommand{\bi}{\begin{itemize}}
\newcommand{\ei}{\end{itemize}}

\newcommand{\bce}{\begin{center}}
\newcommand{\ece}{\end{center}}

\begin{document}

\title{Higher-order exceptional points in a non-reciprocal waveguide beam splitter}

\author{Hamed Ghaemidizicheh,\authormark{1,*} Shahram Dehdashti,\authormark{2} Andreas Hanke,\authormark{1} Ahmed Touhami,\authormark{1} AND Janis N\"{o}tzel,\authormark{2} }

\address{\authormark{1}Department of Physics and Astronomy, University of Texas Rio Grande Valley, Texas 78539, USA\\
\authormark{2}Emmy-Noether Group Theoretical Quantum Systems Design, 
Technical University of Munich, Munich, Germany}

\email{\authormark{*}hamed.ghaemidizicheh@utrgv.edu} 


\begin{abstract*} 
Non-Hermitian systems have attracted significant interest because of their intriguing properties, including exceptional points (EPs), where eigenvalues and the corresponding eigenstates coalesce. In particular, quantum systems with EPs exhibit an enhanced sensitivity to external perturbations, which increases with the order of the EP. Therefore, higher-order EPs hold significant potential for advanced sensing applications, but they are challenging to achieve due to stringent symmetry requirements. In this work, we study the dynamics of a generalized lossy waveguide beam splitter with asymmetric coupling by introducing non-reciprocity as a tunable parameter to achieve higher-order EPs even without dissipation. Moreover, we analyze the evolution of NOON-states under activated non-reciprocity, highlighting its impact on quantum systems. Our results open new pathways for realizing higher-order EPs in non-reciprocal open quantum systems.

\end{abstract*}

\section{Introduction}
In recent years, non-Hermitian systems have attracted significant attention due to their distinct properties such as non-Hermitian edge states \cite{AlvarezPRB2018,Yao2018PRL} and transport effects \cite{ghaemiPRA2021,ghaemiPRB2023}. In particular, non-Hermitian systems are characterized by the existence of exceptional points (EPs), where two or more eigenvalues and their corresponding eigenstates become degenerate
\cite{HeissJOP2004,HeissJOP22012,ding2022non}.
Among the unique properties of non-Hermitian systems due to EPs are topological transport phenomena \cite{xu2016topological,PhysRevLettXu2023,scienceXu2023}, robust wireless power transfer \cite{assawaworrarit2017robust}, asymmetric mode switching \cite{doppler2016dynamically}, stopping light \cite{PhysRevLett.120.013901}, loss-induced suppression and revival of lasing \cite{doi:10.1126/science.1258004,PhysRevA.94.043829}, non-reciprocity \cite{PhysRevA.94.043829}, and enhanced classical sensitivity \cite{PhysRevLett.112.203901,hodaei2017enhanced,chen2017exceptional}.
One of the most important properties of EPs in non-Hermitian systems is the sensitivity to changes in external parameters \cite{hodaei2017enhanced,chen2017exceptional,zhang2018phonon},
which finds important applications in a wide variety of systems \cite{wang2020electromagnetically,xu2016topological,Xiao2023PRB}.
Since the singularity-enhanced sensitivity increases with the order of the EP, i.e., the number of degenerate eigenmodes, systems with higher-order EPs offer even greater promise for developing advanced sensors and precise measurement equipment required in diverse fields \cite{PhysRevLett.127.186601}.
To illustrate the origin of this high sensitivity, consider the eigenvalues of an 
arbitrary non-Hermitian matrix $M(\gamma)$ as a function of a parameter $\gamma\in\CC$ with an $n$-fold EP at $\gamma_{\text{EP}}$. In the vicinity of $\gamma_{\text{EP}}$, different 
eigenvalues $\lambda$ of $M(\gamma)$ are separated by a gap 
$\delta\lambda$ given by a series \cite{Ashida2020book}
\be
\delta\lambda=\sum_{j=1}^{\infty}c_{j}(\gamma-\gamma_{\text{EP}})^{j/n} \, ,
\nonumber\ee
where $c_{j}$ are coefficients. The leading contribution of the
derivative of $\delta\lambda$ with respect to $\gamma$ in the vicinity of the EP is given by
\be
\partial\delta\lambda/\partial\gamma \sim \dfrac{1}{(\gamma-\gamma_{\text{EP}})^{1-1/n}} \, ,\nonumber
\ee
which becomes singular at the EP since $n \ge 2$. This clearly shows a high sensitivity 
of the eigenvalue gap $\delta \lambda$ to the variation of the parameter $\gamma$ away from the EP, which increases with the order $n$ of the EP.

However, realizing higher-order EPs poses practical challenges across different platforms. The primary obstacle originates from the necessary fine-tuning of system parameters, which is required by a higher-order symmetry necessary for eigenmode degeneracy \cite{Teimourpour2018}. To overcome this challenge, waveguides are promising candidates \cite{dehdashti2015realization,setare2019photonic}. For instance, Quiroz-Ju\'{a}rez et al.~realized higher-order EPs in a single, lossy waveguide beam splitter in the quantum domain \cite{Quiroz-Juarez:19}. To this end, they studied
the dynamics of a lossy waveguide beam splitter within the post-selected $N$-photon subspace, where the creation of higher-order NOON-states and number-resolving photon detectors is required instead of fine-tuning. 
In experimental systems, introducing dissipation while maintaining control over other parameters of the beam splitter can be challenging. Furthermore, engineering specific dissipation rates without incurring undesirable losses is difficult. Dissipation can also diminish interference effects, thereby limiting the efficacy of the beam splitter in applications such as interferometry or quantum state manipulation.

To address these problems, in this study we consider a generalized version of a lossy waveguide beam splitter by introducing an asymmetric coupling between its waveguides. In classical systems,
such non-reciprocal platforms have numerous applications including optics and photonics 
\cite{Longhi2015,feng2017non,miri2019exceptional,Ghaemi-Dizicheh:23}, electrical circuits
\cite{helbig2020generalized,zou2021observation,zhao2024exceptional},
optomechanical systems \cite{PhysRevA.111.033512}, 
and nonlinear topological models \cite{Ghaemi-Dizicheh_2024}.
In the quantum regime, non-reciprocal devices, including isolators, circulators, and directional 
amplifiers, offer valuable capabilities for photon- and phonon-based quantum information processing and sensing, as they enhance information throughput while shielding sensitive components from unwanted signal interference \cite{verhagen2017optomechanical}.

Optical non-reciprocity refers to the ability of a system to allow photon transmission in one direction while suppressing it in the opposite direction, thereby breaking time-reversal symmetry. Several mechanisms can induce non-reciprocal behavior in the context of waveguide beam splitters. For example, in a recently proposed approach to designing a non-reciprocal 
waveguide beam splitter \cite{Longhi20}
a lossy waveguide beam splitter is sandwiched between a pair of 50\textbackslash 50 beam splitter which perform forward and backward rotation on input\textbackslash output states. The dynamics of such a rotated system can be governed by a non-reciprocal effective Hamiltonian. This approach has been recently applied experimentally in \cite{sciadv.adr8275}
to investigate the impact of exceptional points on quantum interference of indistinguishable photons.

Non-reciprocal photonic systems have garnered significant attention in recent years, with various practical platforms proposed and experimentally realized. One of the most prominent approaches to achieving non-reciprocal behavior involves time-modulated photonic structures, as extensively discussed in Ref.~\cite{sounas2017non},
where time-dependent refractive index modulation enables direction-dependent transmission. Furthermore, integrated photonic platforms \cite{bi2011chip} 
demonstrate on-chip optical isolation via magneto-optic effects, which can break reciprocity at the material level. 
More recently, waveguide-based non-reciprocity has been explored through evanescent field engineering and spatial asymmetries. 
Notably, non-reciprocity is not merely a design feature but emerges as a fundamental, inherent property in photonic systems, 
particularly in structures with non-identical waveguides. For example, 
Vicencio et al.~recently derived a general formalism for coupling coefficients in 
asymmetric waveguide arrays \cite{vicencio2024non},
showing that non-reciprocal coupling can naturally emerge due to differences in the refractive index profiles of adjacent waveguides, 
an approach that aligns closely with the physical realization of our model.
This insight supports our assertion that non-reciprocity 
can be intrinsic to the physical structure of photonic systems rather than requiring external biasing or modulation.
In light of these developments, we emphasize that our model is consistent with such waveguide-based systems, 
wherein non-reciprocal coupling strengths can be tuned by modifying the local geometry and refractive index contrast between neighboring channels \cite{vicencio2024non}.

In this work, we explore the role of non-reciprocity in engineering EPs within a lossy waveguide beam splitter. By introducing non-reciprocity as an additional degree of freedom, we demonstrate that it facilitates the emergence of higher order exceptional points without stringent dissipation requirements. Our analysis reveals that increasing non-reciprocity lowers the critical dissipation threshold for EP formation, and in the extreme case of a unidirectional system, EPs arise in the absence of dissipation.
The paper is organized as follows. In Section \ref{sec:theory}, we derive the eigenvalues of a non-reciprocal, lossy waveguide beam splitter using the Schwinger representation approach and numerically study properties of the higher-order EPs for this system.
We also demonstrate how altering the system's non-reciprocity affects the
realization of EPs of order $N+1$ within the post-selected $N$-photon subspace. In Section \ref{sec:numeric}, we study the dynamical evolution of a NOON-state both analytically and numerically by activating non-reciprocity in this system.
Section \ref{sec:conclusions} contains our conclusions. 

\section{Spectrum of a Non-Reciprocal, Lossy Waveguide Beam Splitter} \label{sec:theory}
\begin{figure}[htbp]
	\centering\includegraphics[width=0.8\textwidth]{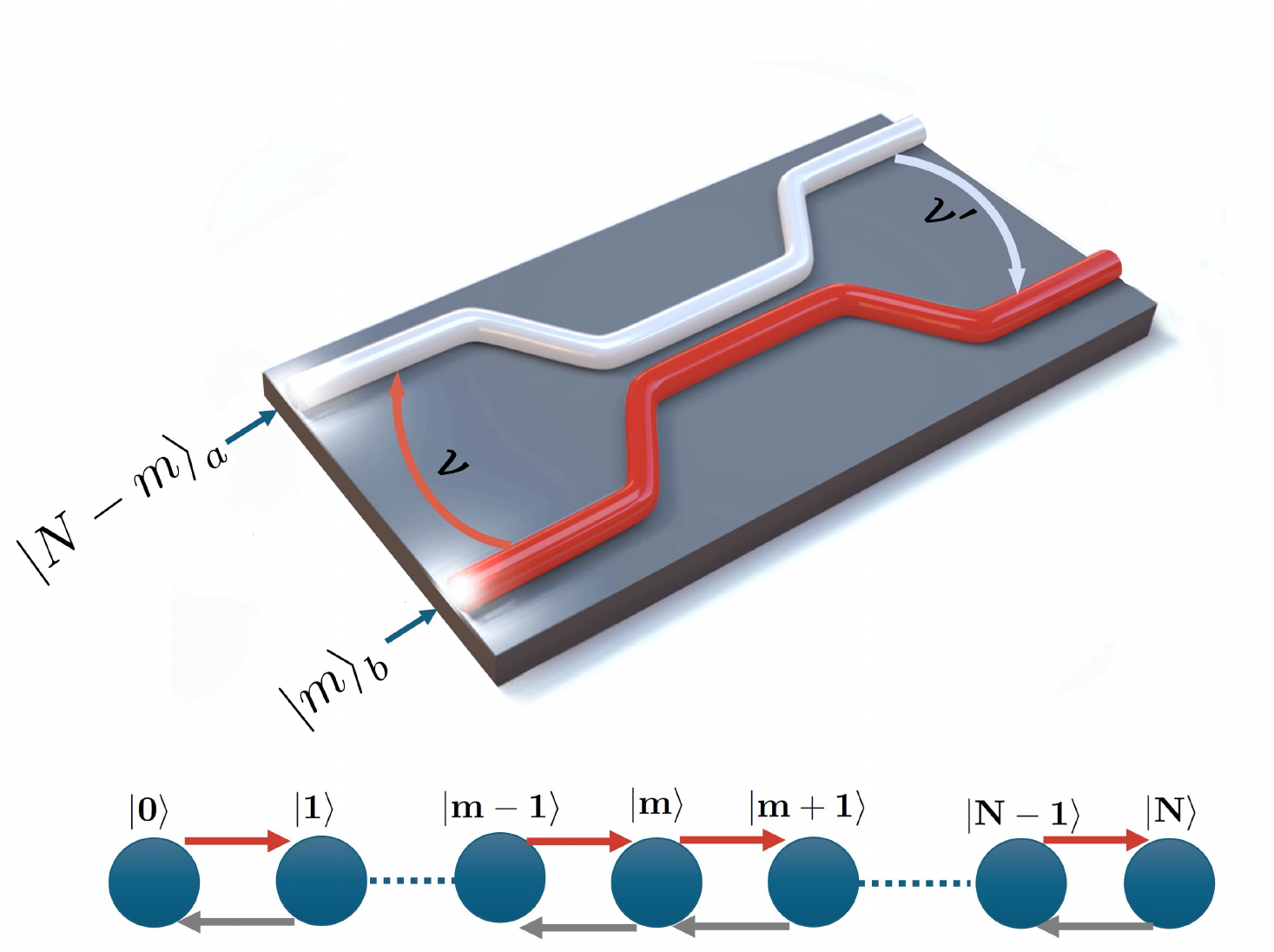}
		\caption{Schematic of a non-reciprocal single, lossy, two-leg waveguide beam splitter excited with $N$ indistinguishable photons prepared in the state 
			$\ket{N-m,m}=\ket{N-m}_a\ket{m}_b$ ($0 \le m \le N)$, where $a$ represents the neutral (white) waveguide and $b$ the lossy (red) waveguide.}
		\label{fig:setup}
	\end{figure}

We consider a non-reciprocal beam splitter with two coupled waveguides depicted schematically in 
Fig.~\ref{fig:setup}. 
The Hamiltonian in second quantization is given by (using units such that $\hbar=1$)
\be
\hat{H}
=\omega_0(\hat{a}^{\dagger}\hat{a}+\hat{b}^{\dagger}\hat{b})+\nu\hat{b}^{\dagger}\hat{a}+\nu'\hat{a}^{\dagger}\hat{b}-i\Gamma \hat{b}^{\dagger}\hat{b} \, ,
\label{nonreciprocal hamiltonian}
\ee
where $\hat{a}^{\dagger}$ $(a)$ and $\hat{b}^{\dagger}$ $(b)$ are bosonic particle creation (annihilation) operators for photonic modes
in the two waveguides and
$\Gamma$ denotes the dissipation coefficient of the lossy waveguide.
The coupling between waveguides $a$ and $b$ is given by $\nu$ and $\nu'$, and $\omega_{0}$ is the on-site energy.
A non-reciprocal waveguide beam splitter is obtained for $\nu \ne \nu'$.
To reveal arbitrary-order EPs in the coupled waveguide beam splitter,
we represent the Hamiltonian in Eq.~\eqref{nonreciprocal hamiltonian} in the two-mode $N$-photon subspace. This subspace is spanned by $N+1$ orthonormal states $\ket{\phi_{m}}:=\ket{N-m,m}=\ket{N-m}_a\ket{m}_b$
where $0 \le m \le N$. 

The Hamiltonian in Eq.~\eqref{nonreciprocal hamiltonian} 
commutes with the total photon-number operator 
\be
\hat{N}=\hat{a}^{\dagger}\hat{a}+\hat{b}^{\dagger}\hat{b} \, ,
\ee
which implies that $N$ is conserved.
Based on the Schwinger representation \cite{Graefe_2008}, 
we introduce ladder operators $\hat{J}_{+}$, $\hat{J}_{-}$ such that
\be
	\hat{J}_{+}=\hat{J}_{x}+i\hat{J}_{y}=\hat{b}^{\dagger}\hat{a} \, , \, \, 
	\hat{J}_{-}=\hat{J}_{x}-i\hat{J}_{y}=\hat{a}^{\dagger}\hat{b} \, ,
\ee
%
which satisfy the $\text{SU}(2)$ commutation relations.
%
%
Expressed in terms of the ladder operators $\hat{J}_{+}$, $\hat{J}_{-}$, the Hamiltonian
in Eq.~\eqref{nonreciprocal hamiltonian} takes the form
\be
\hat{H}=(\omega_0-i\Gamma/2)\hat{N}+\nu \hat{J}_+ +\nu'\hat{J}_--i\Gamma \hat{J}_{z} \, .
\label{Hamiltoniansu(2)}
\ee
The eigenvalues can be readily derived analytically by diagonalizing $\hat{H}$
in Eq.~\eqref{Hamiltoniansu(2)}. To this end, we use
the Baker–Campbell–Hausdorff formula and $\text{SU}(2)$ commutation relations to obtain the following boosts (or rotations):
\bea
&&e^{i\theta \hat{J}_{\pm}}\hat{J}_ze^{-i\theta \hat{J}_{\pm}}=\hat{J}_z\mp i\theta \hat{J}_{\pm} \, , \label{boost1su(2)}\\
&&e^{i\theta \hat{J}_z}\hat{J}_{\pm}e^{-i\theta \hat{J}_{z}}=e^{\pm i\theta}\hat{J}_{\pm} \, , \label{boost2su(2)}\\
&&e^{i\theta \hat{J}_+}\hat{J}_{-}e^{-i\theta \hat{J}_{+}}=\hat{J}_{-}+2i\theta \hat{J}_{z}-(i\theta)^2\hat{J}_{+} \, , \label{boost3su(2)}\\
&&e^{i\theta \hat{J}_-}\hat{J}_{+}e^{-i\theta \hat{J}_{-}}=\hat{J}_{+}-2i\theta \hat{J}_{z}-(i\theta)^2\hat{J}_{-} \, . \, \label{boost4su(2)}
\eea
Since the Casimir operator of the system is given by 
$\hat{C} = (\hat{N}/2) (\hat{N}/2 + 1)$  
it follows that the operator $\hat{N}$ commutes with all the generators of $\text{SU}(2)$. Consequently, using Eqs.~\eqref{boost1su(2)} and \eqref{boost4su(2)}, the Hamiltonian in Eq.~\eqref{Hamiltoniansu(2)} transforms as follows: 
\bea
&&\hat{H}'=e^{i\theta \hat{J}_{-}} {\hat H} e^{-i\theta \hat{J}_{-}}\nonumber\\
&&~~~~=(\omega_0-i\Gamma/2)\hat{N}+\nu \hat{J}_{+}+(\nu\theta^2+\nu'+\Gamma\theta)\hat{J}_{-}\nonumber\\
&&~~~~~~~-i(2\theta\nu+\Gamma)\hat{J}_{z} \, .
\label{ham_transformed_1}
\eea
Setting
\be
\theta=\frac{-\Gamma^{2}\pm\sqrt{\Gamma-4\nu\nu'}}{2\nu}
\ee
eliminates the third term (proportional to $\hat{J}_{-}$) on the right-hand side of Eq.~\eqref{ham_transformed_1}. 
Using Eq.~\eqref{boost1su(2)}, the transformed Hamiltonian 
becomes
\be
\hat{H}'=(\omega_0-i\Gamma/2)\hat{N}-i(2\theta\nu+\Gamma)e^{-i\phi \hat{J}_{+}}\hat{J}_{z}e^{i\phi \hat{J}_{+}} \, ,
\label{ham_transformed_2}
\ee
where
\be
\phi=\frac{\nu}{2\theta\nu+\Gamma} \, .
\ee
By applying the similarity transformation $U\hat{H}'U^{-1}$, where $U=e^{i\phi \hat{J}_{+}}$, we can diagonalize the Hamiltonian $\hat{H}'$. The eigenvalues are then expressed in analytical form as
\be
\lambda_r=(\omega_{0}-i\Gamma/2)N + r\sqrt{4\nu\nu'-\Gamma^{2}} \, , 
\label{eigenvalue}
\ee
where $r=-N/2,-N/2+1,\ldots,N/2$ corresponding to the parameterization of the eigenvalues of an $\text{SU}(2)$ (spin) operator. 
In the reference frame of the associated eigenvectors $\ket{\lambda_{r}}$
the eigenvectors of the Hamiltonian \eqref{nonreciprocal hamiltonian} are given by
$e^{-i\theta} e^{-i\phi} \vec{\nu}_i$ for $i=0, \ldots, N$,
where $\vec{\nu}_i$ is a column vector with components $\nu_{i,k} = \delta_{ik}$,
$k=0, \ldots, N$, where $\delta_{ik}$ is the Kronecker delta.

To study the effect of non-reciprocity, we parametrize the coupling amplitudes $\nu$, $\nu'$ such that
\be
	\nu=\nu_0(1+\eta) \, , \, \,
	\nu'=\nu_0(1-\eta) \, ,
	\label{param_nu}
\ee
where $0 \leq \eta \leq 1$ is the non-reciprocity parameter.
\begin{figure}[htbp]
	\centering\includegraphics[width=1.0\textwidth]{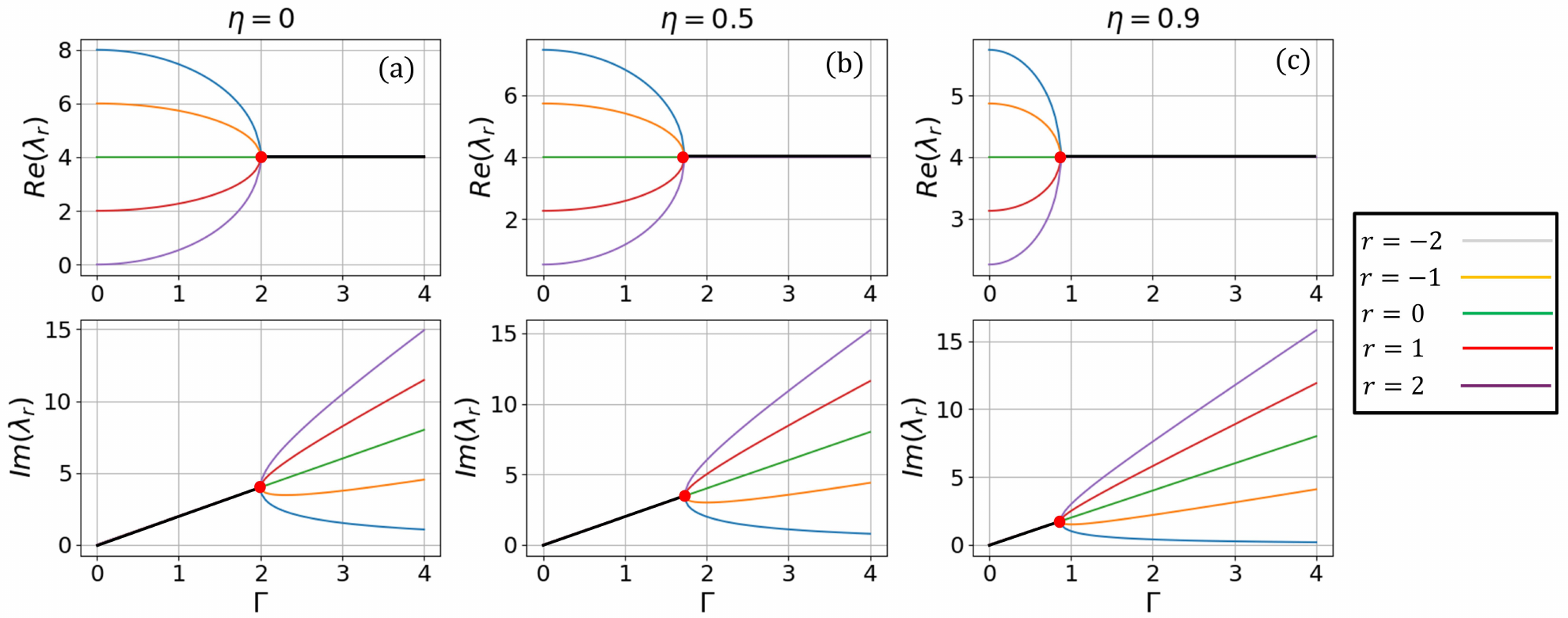}
		\caption{Real and Imaginary parts of the eigenvalues $\lambda_r$ in Eq.~\eqref{eigenvalue}
        of the Hamiltonian in Eq.~\eqref{Hamiltoniansu(2)} (with $\omega_{0}=\nu_{0}=1\,\text{cm}^{-1}$ 
			in Eqs.~\eqref{Hamiltoniansu(2)}, \eqref{param_nu})
			vs.~the dissipation constant $\Gamma$ 
			for $N=4$ with (a) $\eta=0$ (reciprocal), (b) 
            $\eta=0.5$, and (c) $\eta=0.9$.
			The system reaches the exceptional point 
            (EP of order five, red dots) at $\Gamma_{c}=2\nu_{0}(1-\eta^2)^{\frac{1}{2}}$. 
            The index $r$ numbers the eigenvalues of an $\text{SU}(2)$ (spin) operator in an $N+1$ dimensional subspace, see text below Eq.\,\eqref{eigenvalue}.
		} \label{P2}
\end{figure}
By adjusting $\eta$ within the range $0 \leq \eta \leq 1$, the system transitions from a reciprocal configuration at $\eta = 0$ to a fully unidirectional coupled waveguide regime at $\eta = 1$.
According to Eq.~\eqref{eigenvalue}, the adjacent eigenvalue difference, defined as 
$\Delta\lambda\equiv\lambda_r-\lambda_{r-1}$, is given by
\be
\Delta\lambda=\sqrt{4\nu_{0}^2(1-\eta^2)-\Gamma^2} \, ,
\label{delta_ev}
\ee
which is purely real when the
dissipation coefficient satisfies $\Gamma\leq 2\nu_0(1-\eta^2)^{\frac{1}{2}}$, and becomes purely imaginary for $\Gamma > 2\nu_0(1-\eta^2)^{\frac{1}{2}}$. At the critical point,
\begin{equation} \label{critical_point}
\Gamma_c = 2 \nu_0 (1-\eta^2)^{\frac{1}{2}} \, ,
\end{equation}
all eigenvalues become degenerate, and all corresponding
eigenmodes coalesce, resulting in an exceptional point of order $N+1$. In the reciprocal system ($\eta=0$) the system reaches the exceptional point 
for $\Gamma_c = 2\nu_{0}$. To attain the exceptional point by enhancing non-reciprocity, it is necessary to minimize the system's dissipation such that at the unidirectional beam splitter ($\eta=1$), the dissipation is completely eliminated. 

Figure \ref{P2} shows the eigenvalues 
$\lambda_r$ in Eq.~\eqref{eigenvalue}
as a function of the loss parameter $\Gamma$ for $\omega_{0}=\nu_{0}=1\,\text{cm}^{-1}$ and $N=4$. 
In Fig.~\ref{P2}(a) we set the non-reciprocity $\eta=0$, 
representing a reciprocal beam splitter \cite{Quiroz-Juarez:19}. As the dissipation coefficient $\Gamma$ increases, the real parts of the eigenvalues exhibit level attraction and become degenerate at $\Gamma_{c} = 2 \nu_{0}$, remaining constant thereafter. Conversely, the imaginary parts of the eigenvalues are identical and grow linearly with $\Gamma$ for $\Gamma < \Gamma_c$, and become non-degenerate for
$\Gamma > \Gamma_c$ \cite{Quiroz-Juarez:19}. 
As shown in Figs.~\ref{P2}(b) and \ref{P2}(c),
increasing the non-reciprocity parameter $\eta$ 
reduces the critical dissipation value $\Gamma_c$ required to reach an exceptional point,
as predicted analytically (Eqs.~\eqref{eigenvalue}~-~\eqref{delta_ev}).
For a unidirectional beam splitter ($\eta=1$) the real parts of the eigenvalues remain uniform and unaffected by dissipation, whereas the imaginary parts increase linearly with $\Gamma$.

\begin{figure}[htbp]
	\centering\includegraphics[width=1.0\textwidth]{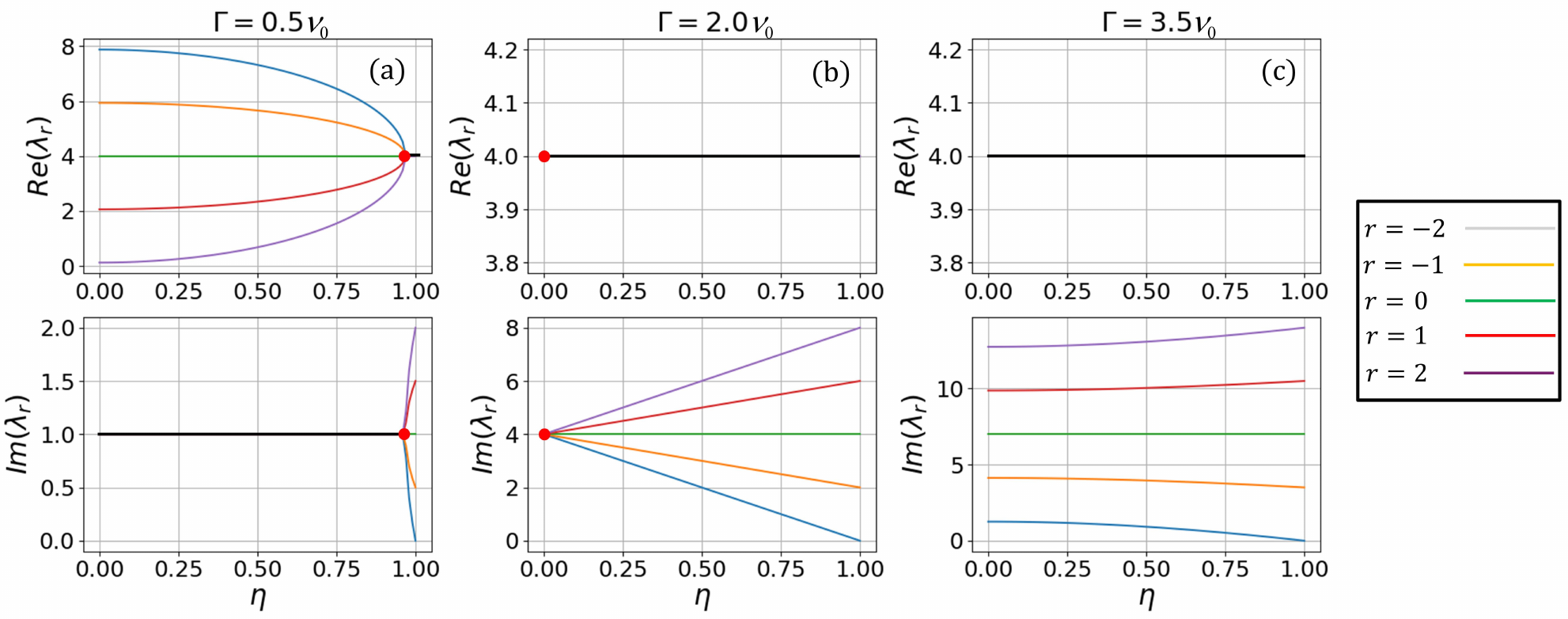}
		\caption{Flow of eigenvalues $\lambda_r$ in Eq.~\eqref{eigenvalue}
        of the Hamiltonian in Eq.~\eqref{Hamiltoniansu(2)} for $N=4$ as a function of the
        non-reciprocity parameter $\eta$ with dissipation values
        (a) $\Gamma=0.5\,\nu_0$, (b) $\Gamma = 2.0\,\nu_0$ (where $\eta_c = 0$),
        and (c) $\Gamma=3.5\,\nu_{0}$. 
        For $\Gamma \in [0,2\nu_{0}]$
        the system reaches the higher-order exceptional point 
        (EP of order five, red dots in (a), (b)) by adjusting $\eta$ 
        to the critical value $\eta_{c}=\sqrt{4\nu_{0}^{2}-\Gamma^{2}} \Big/ (2\nu_{0})$, 
        whereas the system does not exhibit an EP for $\Gamma > 2 \nu_0$ 
        (as shown in (c)).
        The index $r$ numbers the eigenvalues of an $\text{SU}(2)$ (spin) operator in an $N+1$ dimensional subspace, see text below Eq.\,\eqref{eigenvalue}.
        }
		\label{P3}
\end{figure}

The non-reciprocal waveguide beam splitter offers an additional degree of flexibility for achieving higher-order exceptional points by modulating the non-reciprocity across the system. For any dissipation value within the range $\Gamma \in [0,2\nu_{0}]$, higher-order exceptional points occur when the non-reciprocity parameter $\eta$ is set to $\eta_{c}=\sqrt{4\nu_{0}^{2}-\Gamma^{2}} \Big/ (2\nu_{0})$. 
Figure \ref{P3} illustrates the flow of eigenvalues
$\lambda_r$ in Eq.~\eqref{eigenvalue}
as a function of the non-reciprocity parameter $\eta$
for different dissipation values, namely 
$\Gamma/\nu_0 = 0.5$, $2.0$, and $3.5$ 
in Figs.~\ref{P3}(a)~-~\ref{P3}(c), respectively.

\begin{figure}[htbp]
	\centering\includegraphics[width=1.0\textwidth]{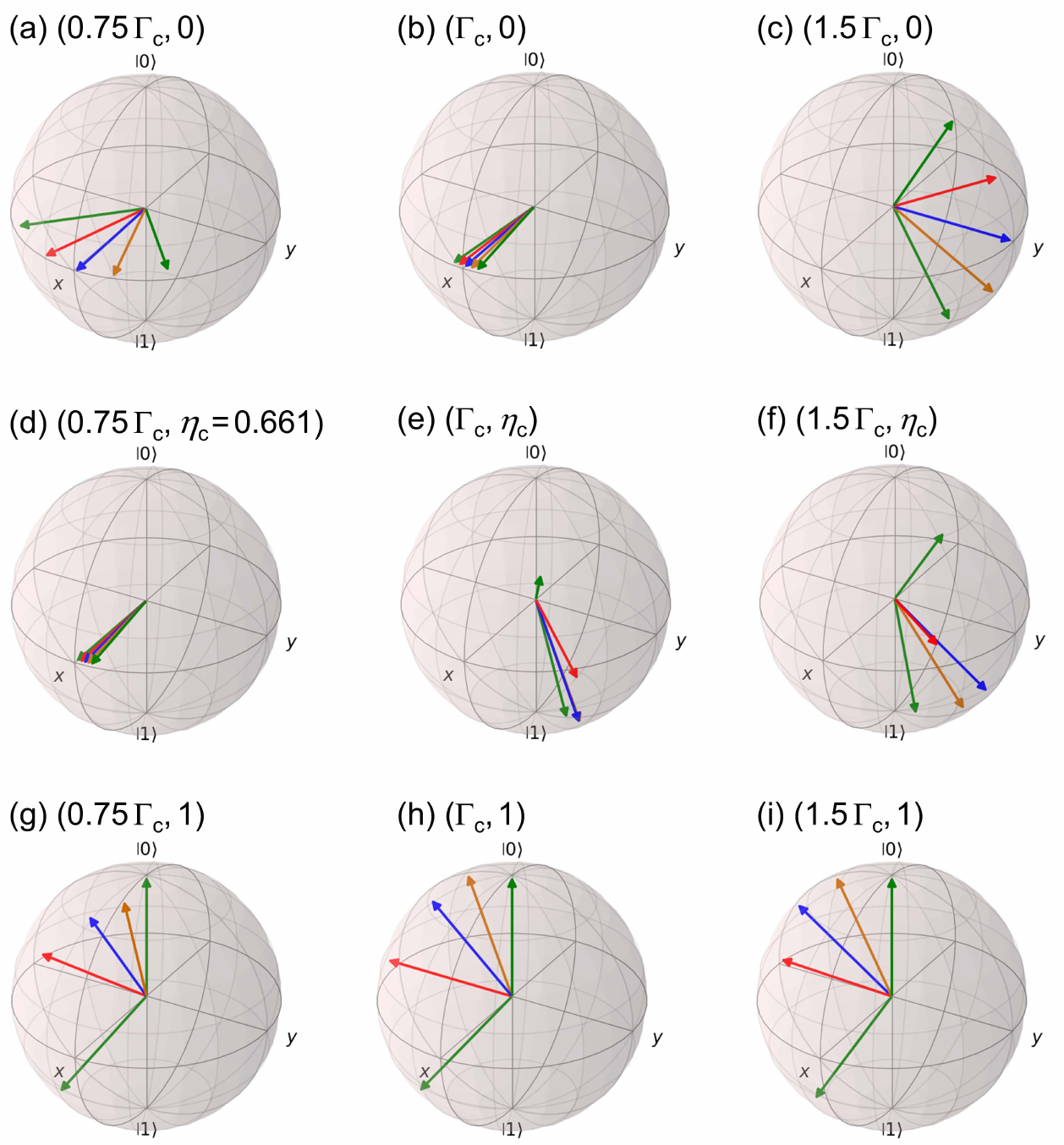}
		\caption{Progression of the spin-projection components $\hat{J}_{x,y,z}$ for $N+1$ eigenmodes $\ket{\lambda_{r}}$
        for different values of the dissipation $\Gamma$ and non-reciprocity $\eta$ as
        indicated in the format $(\Gamma, \eta)$. Here we consider $N=4$ photons.}
		\label{P4}
\end{figure}

Figure \ref{P4} provides a visual representation of the effect of the 
loss parameter $\Gamma$ and non-reciprocity parameter $\eta$ 
on the exceptional point by calculating the
expectation values of  three spin-projection components  
$J_{\alpha, r} = \bra{\lambda_r } \hat{J}_\alpha \ket{ \lambda_r}$,
$\alpha \in \{x, y, z\}$, where $\ket{\lambda_{r}}$ is a complex
eigenvector of the Hamiltonian \eqref{Hamiltoniansu(2)}. Figure \ref{P4} shows the evolution of these components for varying $\Gamma$, $\eta$
for $N=4$ photon subspaces. Each subplot represents a distinct combination of
these parameters to demonstrate their effect on the eigenmodes. 
In Figs.~\ref{P4}(a)~-~\ref{P4}(c) (top row), 
the system remains reciprocal ($\eta=0$), showing how increasing $\Gamma$ affects the spin projections. 
At $\Gamma=0.75\,\Gamma_c$, the spin projections lie in the $x-y$ plane
(Fig.~\ref{P4}(a));
if $\Gamma$ reaches the exceptional point $\Gamma_{c}$, the eigenvectors 
coalesce, signifying the system's transition through the exceptional point
(Fig.~\ref{P4}(b)).
As dissipation increases beyond this critical value, 
the eigenvalues are located in the $y-z$ plane
(Fig.~\ref{P4}(c)).

In Figs.~\ref{P4}(d)~-~\ref{P4}(f) (middle row),
the system incorporates intermediate non-reciprocity $\eta=\eta_c=0.661$. 
Notably, in Fig.~\ref{P4}(d)
the system reaches the exceptional point at a lower dissipation value $\Gamma=0.75\,\Gamma_c$, 
indicating that the presence of non-reciprocity allows the system to transition through the 
exceptional point earlier than in the reciprocal case. 
This demonstrates how non-reciprocity effectively lowers the threshold for reaching the exceptional point.

In Figs.~\ref{P4}(g)~-~\ref{P4}(i) (bottom row),
the system is fully unidirectional ($\eta=1$), hence non-reciprocity dominates the system dynamics. Unlike the previous cases, where the system transitions to the exceptional point at specific loss values, here the system has already reached the exceptional point in a lossless regime ($\Gamma=0$). Consequently, the variations in dissipation beyond this point no longer induce a significant structural change in the spin-projections. This is reflected by the fact that the three plots shown in 
Figs.~\ref{P4}(g)~-~\ref{P4}(i)
appear nearly identical, indicating that the spin projections remain robust even as dissipation increases. This confirms that in a unidirectional system, the exceptional point is dictated primarily by non-reciprocity rather than loss, leading to a saturation effect where further dissipation does not significantly alter the eigenmode structure.

\section{Dynamics of the non-reciprocal beam splitter} \label{sec:numeric}

In this section, we explore the dynamics of the lossy non-reciprocal beam splitter. 
The goal is to find the behavior of the intensity $I(z)$ within the $N$-photon subspace as a function 
of the propagation distance $z$ in the waveguide. Since the waveguide is homogeneous 
in $z$-direction, and since we use units such that the propagation velocity 
$v$ in the waveguide is $v=1$, the behavior $I(z)$ directly corresponds to the time evolution of the intensity $I(t)$
as a function of time $t$ at a fixed position in the waveguide.
To obtain a photon-number-resolved population, we need to find the time evolution operator $\hat{\cG}(z)$ for a 2-leg lossy beam splitter
satisfying the Schr\"odinger equation
{\color{black}(identifying the propagation distance $z$ in the waveguide with time, 
	see the remark above)}
\be
i\partial_z\hat{\cG}(z)=\hat{H}\hat{\cG}(z) \, . 
\label{se}
\ee
To find the solution of Eq.~\eqref{se} we use the Wei-Norman method \cite{wei1963lie,Charzynski_2013} to express the time evolution in terms of $\text{SU}(2)$ operators $\hat{J}_{\pm}$, $\hat{J}_{z}$ and the total number operator $\hat{N}$. It is given by
\be
\hat{\cG}(z)=e^{-i(\omega_{0}-i\Gamma/2)\hat{N}z}e^{-if_{+}(z)\hat{J}_{+}}e^{-if_{z}(z)\hat{J}_{z}}e^{-if_{-}(z)\hat{J}_{-}}.
\label{te}
\ee
By substituting the time-evolution operator $\hat{\cG}(z)$ into the Schr\"{o}dinger equation, one can show that the three complex-valued functions $f_{\pm}(z)$ and $f_{z}(z)$ satisfy the following set of coupled, nonlinear differential equations:
\bea
&&\partial_{z}f_{-}=\nu'e^{-if_{z}} \, ,\label{fminus}\\
&&\partial_{z}f_{+}=\nu+\nu'f_{+}^2-\Gamma f_{+} \, ,\label{fplus}\\
&&\partial_{z}f_{z}=-i\Gamma+2i\nu'f_{+} \, .\label{fz}
\eea
The analytical solutions to these three nonlinear equations for the initial condition
\( G(0) = 0 \), equivalent to \( f_{\pm}(0) = f_z(0) = 0 \), are expressed as follows:
\bea
&&f_z(z)=-2i\ln\left[\cos(z\Delta\lambda/2)+\frac{\Gamma}{\Delta\lambda}\sin(z\Delta\lambda/2) \right], \nonumber\\
\nonumber\\
&&f_{+}(z)=\frac{\Gamma}{2\nu'}+\frac{\Delta\lambda}{2\nu'}\left[ \frac{\tan(z\Delta\lambda/2)-\frac{\Gamma}{\Delta\lambda}}{1+\frac{\Gamma}{\Delta\lambda}\tan(z\Delta\lambda/2)}\right],\nonumber \\
\nonumber\\
&&f_{-}(z)=\frac{\Gamma}{2\nu}+\frac{\Delta\lambda}{2\nu}\left[ \frac{\tan(z\Delta\lambda/2)-\frac{\Gamma}{\Delta\lambda}}{1+\frac{\Gamma}{\Delta\lambda}\tan(z\Delta\lambda/2)}\right].
\eea
The functions $f_{\pm}$ are real, irrespective of whether the system passes through an exceptional point, but $f_{z}(z)$ is complex in general. Note that unlike the case of a reciprocal beam splitter, here $f_{+}(z) \neq f_{-}(z)$, reflecting the non-reciprocal nature of the system.

If the system approaches a unidirectional regime, i.e., $1-\eta := \epsilon\ll 1$, the $f$-functions can be expanded as
\bea
&&f_z(z)=-i\Gamma z-\frac{4i\nu_0^2(1-e^{-\Gamma z}-z\Gamma)\epsilon}{\Gamma^{2}}+\cO(\epsilon^2) \, ,\nonumber\\
&&f_{+}(z)=\frac{2\nu_0^3}{\Gamma}(1-e^{-\Gamma z})\epsilon^{2}+\cO(\epsilon^3) \, ,\nonumber\\
&&f_{-}(z)=\frac{4\nu_0^3}{\Gamma}(1-e^{-\Gamma z})\epsilon+\cO(\epsilon^2) \, .
\eea
Thus, for $\epsilon \to 0$, corresponding to the unidirectional limit $\eta \to 1$,
$f_{+}$ approaches zero more rapidly than the other $f$-functions. When the system reaches the unidirectional regime, the only significant contribution to the time evolution is $f_{z}(z)$, which simplifies to a linear dependence on the dissipation rate.
\begin{figure}[htbp]
	\centering\includegraphics[width=1.0\textwidth]{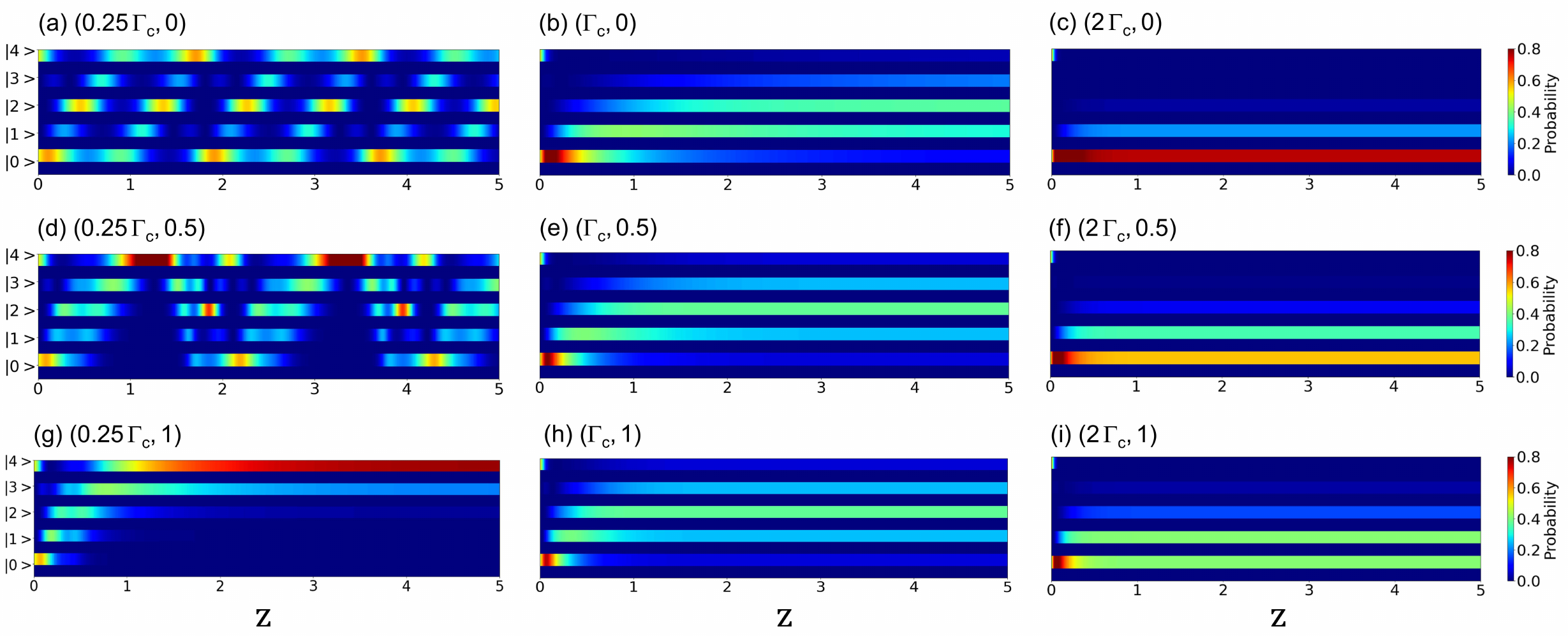}
		\caption{Mode occupation dynamics 
        for $N=4$ photons in terms of the $z$-dependent occupation function $P(m;\ket{\psi(z)})$ in Eq.~\eqref{occfun} for different values of the dissipation $\Gamma$ and non-reciprocity $\eta$,
        indicated in the format $(\Gamma, \eta)$.
        The mode numbers $m = 0, \ldots, 4$ are shown in the vertical bar on the left.}   
		\label{P5}
\end{figure}

To investigate the dynamics of the non-reciprocal waveguide beam splitter, we examine the evolution of a NOON-state, defined as the initial state
\begin{equation}\label{noon}
\ket{\text{NOON}}=
\ket{\psi(0)}= \frac{1}{\sqrt{2}} 
\left( \ket{N}_{a}\ket{0}_b+\ket{0}_{a}\ket{N}_{b} \right) \, ,
\end{equation}
where $\ket{N}_a$ and $\ket{N}_b$ represent the $N$-photon states in the 
respective modes $a$ and $b$. 
The choice of the NOON state as the initial input is motivated by several key considerations. First, the Casimir operator, which commutes with the system’s Hamiltonian, is proportional to the total excitation number operator. Consequently, our analysis is naturally confined to subspaces with a fixed total number of photons. Within these subspaces,
the NOON-state stands out as a highly symmetric, maximally path-entangled quantum state that exhibits pronounced quantum interference effects \cite{science.1188172}.
Moreover, the NOON-state is well known for its exceptional sensitivity to relative phase shifts, making it a valuable resource in quantum metrology and quantum information protocols 
\cite{Israel2014,Dowling01032008,Giovannetti2011}.
In the context of our non-reciprocal waveguide beam splitter, it serves as an ideal probe for detecting direction-dependent asymmetries and enhanced sensitivity to non-Hermitian dynamics. Its evolution under non-reciprocal coupling highlights interference features and signatures of non-reciprocity that would be less prominent with separable or weakly entangled states.

To analyze the system's evolution, our numerical approach involves post-selecting a subset of the system's state space (manifold) corresponding to a specific quantum outcome. Specifically, for an $N$-photon input, we consider only those cases where the total number of photons detected is exactly $N$ and no photons are absorbed in the lossy waveguide.
We consider the $z$-dependent occupation function 
\begin{eqnarray} \label{occfun}
	P(m;\ket{\psi(z)}) = \frac{|\langle m | \psi(z) \rangle|^2}{\braket{\psi(z) | \psi(z)}} \, ,
\end{eqnarray}
which is normalized such that $\sum_{m=0}^{N}P(m;\ket{\psi(z)})=1$.
\begin{figure}[htbp]
	\centering\includegraphics[width=1.0\textwidth]{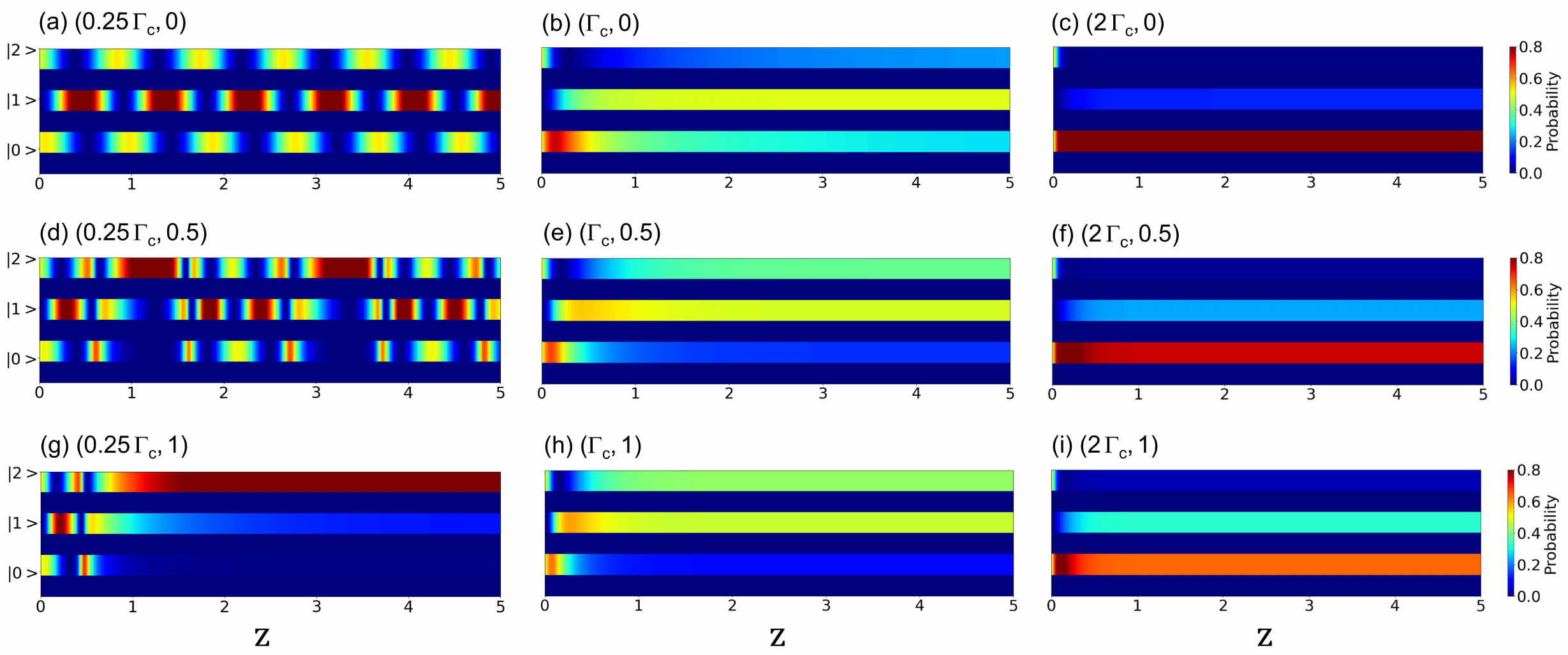}
		\caption{Mode occupation dynamics for $N=2$ photons
        in terms of the $z$-dependent occupation function $P(m;\ket{\psi(z)})$ in Eq.~\eqref{occfun}
        for different values of the dissipation $\Gamma$ and non-reciprocity $\eta$,
        indicated in the format $(\Gamma, \eta)$.
        The mode numbers $m = 0, 1, 2$ are shown in the vertical bar on the left.
        The values of $\Gamma$ and $\eta$ in panels (a) - (i) are the same as in Fig.~\ref{P5}.}   
		\label{P6}
\end{figure}
Figure \ref{P5} shows $P(m;\ket{\psi(z)})$
in Eq.~\eqref{occfun} as a function of the propagation distance $z$ for $N=4$ photons
(corresponding to the time evolution of the mode occupation at a fixed position 
in the waveguide, see text above Eq.~\eqref{se})
for different values of the loss parameter $\Gamma$ 
and non-reciprocity parameter $\eta$. 
In the reciprocal regime, when the dissipation is below the exceptional point $\Gamma_c$, the mode occupation exhibits oscillatory behavior across the five modes, with energy transferring from regions of lowest occupation to highest occupation 
(Fig.~\ref{P5}(a)).
At the exceptional point $\Gamma_c$, the mode occupation stabilizes, with energy accumulating in the low-loss region 
(Fig.~\ref{P5}(b)). This behavior persists when dissipation exceeds the critical value $\Gamma_c$ 
(Fig.~\ref{P5}(c)).
In the non-reciprocal regime and for low loss, where $\eta \neq 0$ and $\Gamma < \Gamma_c$, the periodic pattern is disrupted, and photons preferentially occupy higher-order modes 
(Figs.~\ref{P5}(d) and \ref{P5}(g)). However, in the non-reciprocal regime where $\Gamma \geq \Gamma_c$, the non-reciprocity parameter no longer plays a significant role, and photons do not occupy higher modes, as shown in
Figs.~\ref{P5}(e), \ref{P5}(f), \ref{P5}(h), and \ref{P5}(i). 
These results emphasize the impact of nonreciprocity on the mode occupation dynamics, particularly in the low-dissipation regime. Beyond the exceptional point, however, the dispersion parameter \(\Gamma\) becomes dominant, effectively stabilizing the lower modes. 

\begin{figure}[htbp]
	\centering\includegraphics[width=0.8\textwidth]{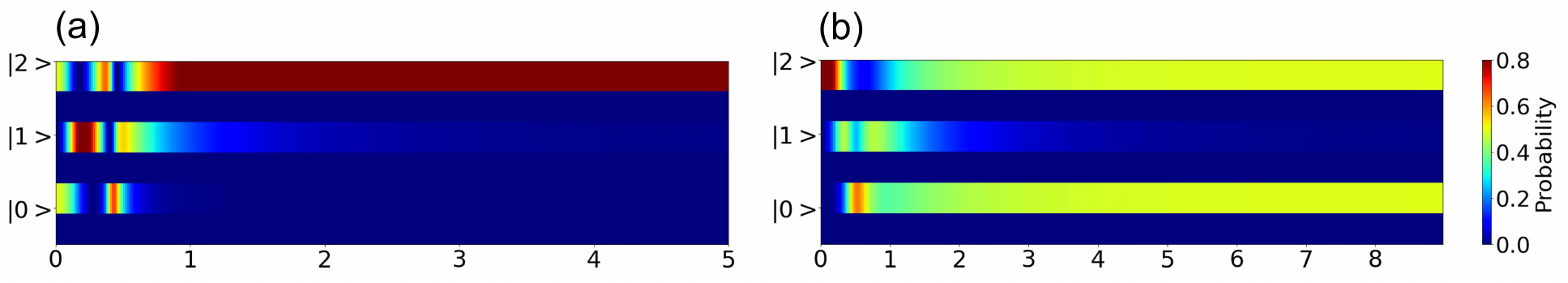}
		\caption{Mode occupation dynamics for $N=2$ photons with $\Gamma=0$, $\eta=1$,
        illustrated via the $z$-dependent occupation probability function
        $P(m; \ket{\psi(z)})$ in Eq.~\eqref{occfun}, for the initial (a) quantum and (b) classical states defined in 
        Eqs.~\eqref{noon} and \eqref{classical}, respectively.
        The parameter values $\Gamma=0$, $\eta=1$ describe a fully unidirectional system without dissipation, 
        corresponding to an exceptional point (see Eq.\,\eqref{critical_point}).} \label{P5and6extend}
\end{figure}

Figure \ref{P6} 
shows the mode occupation dynamics in terms of $P(m;\ket{\psi(z)})$ in Eq.~\eqref{occfun} 
for $N=2$ photons; the values of the loss parameter $\Gamma$ and non-reciprocity parameter $\eta$ in 
Figs.~\ref{P6}(a)~-~\ref{P6}(i) are the same as in Fig.~\ref{P5}. 
The behavior of photon occupation follows a similar pattern to that observed in Fig.~\ref{P5} for $N=4$. Specifically, as the loss parameter increases from the left and surpasses the exceptional point, the photon occupation stabilizes, resembling the trends shown in Fig.~\ref{P5}. This stability occurs as the dissipation exceeds the exceptional point, leading to a stable energy distribution across the modes, analogous to the behavior seen in the corresponding plots for $N=4$. Although the overall mode occupation dynamics for $N=2$ resembles those observed for $N=4$, the two-photon case holds particular importance due to the Hong-Ou-Mandel (HOM) effect \cite{PhysRevLett.59.2044, Lewis-Swan2016,jaouni2025tutorial}. In Hermitian systems, the HOM effect manifests itself as photon bunching, where two identical photons interfere destructively in certain modes, leading to their coalescence in the same output state. Notably, as indicated in Fig.~\ref{P5and6extend}(a), in the \
low-dissipation regime the HOM effect is observed. 
However, when the dissipation is equal to or greater than the exceptional point the HOM effect is no longer achievable.

To highlight the distinct nature of quantum dynamics in our non-reciprocal beam splitter model, 
in Fig.~\ref{P5and6extend} we compare the evolution of a quantum-coherent NOON-state in Eq.~\eqref{noon} with a classical-like statistical mixture described by the density operator
\begin{equation}\label{classical}
\hat{\rho}_{\text{cl}} = \frac{1}{2} 
\left( \ket{N,0} \hspace{-2pt} \bra{N,0} + \ket{0,N} \hspace{-2pt} \bra{0,N} \right) \, .
\end{equation}
The latter captures an incoherent probabilistic distribution of $N$ photons between the two waveguides, corresponding to classical-light coupling into either port, without phase coherence.
The parameter values $(\Gamma=0, \eta=1)$ used in Fig.~\ref{P5and6extend}
describe a fully unidirectional system without dissipation, corresponding to an exceptional point 
(see Eq.~\eqref{critical_point}).
As shown in Fig.~\ref{P5and6extend}(a), for the quantum-coherent 
NOON-state in Eq.~\eqref{noon} photons predominantly occupy the highest Fock state channel 
(mode number $m=2$), revealing direction-dependent asymmetries in the population dynamics. 
This quantum state exhibits HOM interference and displays clear signatures of interference 
fringes and amplitude revivals.
In contrast, for the classical statistical mixture in Eq.~\eqref{classical}
shown in Fig.~\ref{P5and6extend}(b), these features are notably absent.
This distinction emphasizes the essential role of quantum coherence in producing non-classical dynamical behavior.
These differences are not only of fundamental interest but also have practical implications: the sensitivity of quantum interference patterns to system parameters can be exploited for high-precision sensing, noise-resilient state discrimination, and programmable quantum routing protocols. This comparison emphasizes the potential utility of coherent quantum inputs in photonic non-reciprocal systems
and highlights how non-reciprocal quantum dynamics offers operational advantages over their classical counterparts.

\section{Conclusions} \label{sec:conclusions}

In this study, we have examined the eigenvalue spectrum and dynamics of a non-reciprocal, lossy waveguide beam splitter to uncover the mechanisms leading to higher-order exceptional points (EPs). By incorporating non-reciprocity as an additional degree of freedom, we demonstrated the feasibility of achieving $(N+1)$-fold exceptional points without stringent requirements for dissipation.

Our analysis illustrates that non-reciprocity significantly influences the eigenvalue spectrum, with the critical dissipation required for EPs decreasing as non-reciprocity increases. In the limiting case of a unidirectional system, exceptional points emerge even in the absence of dissipation, highlighting the potential of non-reciprocal waveguide beam splitters for robust sensing and precision measurement applications. Furthermore, our results for the eigenvector evolution offer valuable insight into the geometric and dynamical characteristics of systems near EPs, with spin projection components serving as an intuitive visualization of coalescing eigenstates.
The results of this work emphasize the versatility of non-reciprocal platforms in engineering higher-order EPs, which hold promise for advanced interferometric systems \cite{PhysRevA.110.033521}, quantum state manipulation \cite{doi:10.1126/sciadv.adi0732}, and next-generation sensing technologies. Future research may extend this framework to nonlinear or time-dependent systems, potentially uncovering new regimes of EP-driven phenomena and expanding the applicability of non-Hermitian physics.
\begin{backmatter}
\bmsection{Funding}
National Science Foundation (PHY-2012172, OSI-2231387); Bayerische Staatsministerium für Wirtschaft, Landesentwicklung und Energie (6GQT); Bundesministerium für Bildung und Forschung (16KISK002, 16KIS1598K, 16KISQ039, 16KISQ077, 16KISQ093); Deutsche Forschungsgemeinschaft (1129/2-1).
\bmsection{Acknowledgment} This work was supported by NSF grants PHY-2012172 and OSI-2231387.\\
This research is part of the Munich Quantum Valley, which is supported by the Bavarian state government with funds from the Hightech Agenda Bayern Plus and received support from the Bavarian Ministry for Economic Affairs (StMWi) via the project 6GQT.
The authors acknowledge the financial support by the Federal Ministry of Education and Research of Germany in the programme of “Souverän. Digital. Vernetzt.” for the Joint project 6G-life, project identification number: 16KISK002 and via grants 16KIS1598K, 16KISQ039, 16KISQ077, 16KISQ093 and that of the DFG via grant 1129/2-1. 
\bmsection{Disclosures}
The authors declare no conflicts of interest.

\end{backmatter}
\bibliography{bibliography}






\end{document}